\documentclass[twocolumn]{aa}
\usepackage{graphicx}
\usepackage{txfonts}
\begin{document}
\title{Evidence for TeV gamma ray emission from TeV~J2032+4130 in 
Whipple archival data}

\author{M. J. Lang
\inst{1}
\thanks{\email{mark.lang@nuigalway.ie}}
\and D. A. Carter-Lewis\inst{2}
\and D.J. Fegan\inst{3}
\and S.J. Fegan\inst{4}
\and A. M. Hillas\inst{5}
\and R. C. Lamb\inst{6}
\and M. Punch\inst{7}
\and P.~T.~Reynolds\inst{8}
\and T.~C.~Weekes\inst{4}
}

\institute{Department of Physics, National University of Ireland,
Galway, Ireland
\and Department of Physics and Astronomy, Iowa State University,
Ames, IA 50011, USA 
\and Department of Experimental Physics, University College, Belfield,
Dublin 4, Ireland 
\and Fred Lawrence Whipple Observatory, Harvard-Smithsonian CfA,
Amado, AZ 85645, USA
\and Department of Physics and Astronomy, University of Leeds, 
Leeds, LS2 9JT, UK
\and Space Radiation Laboratory, California Institute of
Technology, Pasadena, CA 91125, USA 
\and Physique Corpusculaire et Cosmologie, Coll\`ege de France,
11 place Marcelin Berthelot, 75231 Paris Cedex 05, France 
\and Department of Applied Physics and Instrumentation,
Cork Institute of Technology, Bishopstown, Cork, Ireland 
}

\date{Received 5 April 2004 / Accepted 12 May 2004}

\abstract{A reanalysis of data taken on the Cygnus region in 1989-90 using the 
Whipple Observatory atmospheric Cherenkov imaging telescope confirms the 
existence of the TeV~J2032+4130 source reported by the Crimean Astrophysical 
Observatory and published by the HEGRA Collaboration. 
The significance at the a priori HEGRA position is 3.3$\sigma$. 
The peak signal was found at RA~=~ 20h32m, Dec~=~+41$^\circ$33'.
This is 0.6$^\circ$ north of Cygnus~X-3 which was the original target of the 
observations.
The flux level (12\% of the level of the Crab Nebula) is intermediate between 
the two later observations and suggests that the TeV source is 
variable.
\keywords{gamma rays:observations}
}
  
\maketitle
%

\section{Introduction}
The most powerful technique for the discovery of cosmic sources of TeV 
gamma rays is the atmospheric Cherenkov technique using cameras consisting 
of arrays of photomultiplier tubes coupled to large optical reflectors. Such
instruments have fields of view that are typically less than 5$^\circ$ and
hence are not suitable for sky surveys. The usual mode of operation is to 
make directed observations of candidate objects (chosen from their properties 
at longer wavelengths) and more than a dozen sources have been thus established
(Horan \& Weekes \cite{horan04}). Sky surveys at these energies are 
comparatively insensitive and have not resulted in the detection of any new 
sources (Weekes, Helmken \& Horine \cite{weekes79}, Atkins et al. 
\cite{atkins04}, Cui, Yan et al. \cite{cui03}).
By contrast the 100 MeV space telescopes are ideally suited for sky surveys 
and have discovered some 270 discrete sources; more than half the sources 
detected by the EGRET telescope on the Compton Gamma Ray Observatory are 
unidentified (Hartman et al. \cite{hartman99}).

The increasing sensitivity of the ground-based gamma-ray telescopes and the 
ability to make two-dimensional maps of the field of view encompassed by
the camera makes possible serendipitous detection of TeV gamma-ray
sources in the vicinity of candidate objects or in the control regions
commonly used in making such observations. Regions of the sky that include
the Galactic Plane are particularly attractive for such searches. Early
indications that Cygnus X-3 might be a TeV source (Vladimirsky, Stepanian \& 
Fomin \cite{vlad73}), led to a concentration of observations in that crowded 
region of the sky. 
Recent observations have not confirmed emission from Cygnus X-3 
but a two-dimensional study by the Crimean group of observations taken in 1993
suggested the existence of new source offset by 0.7$^\circ$ from Cygnus
X-3 (Neshpor et al. \cite{neshpor95}). The signal was at the 6$\sigma$ level 
before taking trial factors into account. 
The observations were made over a six week period
and correspond to a flux of 3~x~10$^{-11}$ photons cm$^{-2}$ s$^{-1}$
above an energy of 1 TeV; this is strong source at approximately the level
of the Crab Nebula, the strongest known steady source in the sky.
The coordinates of the new source were Right 
Ascension = 20h~32m and Declination = +41$^\circ$~37'. The uncertainty
in the source position was estimated as 0.2$^\circ$. Subsequent
observations by the Crimean group in the following years did not
confirm the emission at this high level (Fomin \cite{fominpriv}).

In a series of observations by the HEGRA group (Aharonian et al. 
\cite{aharonian02}), the source was independently and serendipitously detected 
in observations of Cygnus X-3 and the EGRET unidentified source, 
GeV J2035+4214 (Lamb \& Macomb \cite{lamb97}).
Both sets of observations encompassed the position of the Crimean source. 
A detection was reported based on 113 hours of observation over a three year 
observing campaign  (1999-2001 inclusive) at the 4.6$\sigma$ post-trial level 
of significance. This large database was expanded to 279 hours with additional 
observations taken in 2002 and the statistical significance was increased to 
the 7$\sigma$ level (Rowell, Horns et al. \cite{rowell03}). 
The source location was determined to be Right Ascension (J2000) = 20h 31m 
57.0s, Declination (J2000) = +41$^\circ$ 29' 56.8" with statistical
and systematic uncertainties $\approx$ a few minutes of arc. This position is 
consistent with that of the Crimean source. 
However the flux level was only about 3\% of the Crab flux 
(F($>$1TeV) = (5.9$\pm$3.1) x10$^{-13}$ photons cm$^{-2}$ s$^{-1}$) and the 
differential spectral index was hard (-1.9).
The source was reported to be extended (6.2') with a point source hypothesis
rejected at the 3$\sigma$ level. 
However, this is a complicated region of the Galactic Plane with three EGRET 
sources nearby but not obviously related. The source lies close to the edge of 
the dense OB association, Cygnus OB2.

Following these reports we were motivated to reanalyze data taken
on Cygnus X-3 with the Whipple telescope at an earlier epoch and at a 
somewhat lower energy threshold to search for evidence of this source.

\section{Observations}

Cygnus X-3 was observed using the Whipple Observatory 10m imaging telescope 
during the epoch 1988-1990, (O'Flaherty et al. \cite{o'flaherty92}).
The observations were carried out in the ON/OFF tracking mode, with Cygnus X-3 
located at the centre of the field of view. In this mode the telescope tracks 
the object for 38 minutes, recording an ON scan. Then during a 2 minute 
interval the telescope is slewed to a position 40 minutes later in Right 
Ascension and records a comparison OFF scan over the same range of zenith 
and azimuth angles.
The telescope was equipped with a 109-pixel camera, consisting of 91 
phototubes of diameter 2.9 cm (with a pixel separation of 
0.25$^\circ$), surrounded by an outer ring of 18 phototubes with a diameter of 
5 cm (with a pixel separation of 0.5$^\circ$), 
(Cawley et al. \cite{cawley90}). 
The field of view of the camera was 3.5$^\circ$ which includes the reported 
location of TeV~J2032+4130. 
The signal in each phototube was amplified and converted to a digital value 
in which each digital count represented a signal of 1.15 photoelectrons.
The camera was triggered when the signal in 2 out of the inner 91 phototubes
exceeded a threshold of about 40 digital counts.
Moment-fitting parameters were used to characterise the shape and orientation 
of each recorded Cherenkov image. Gamma-ray-like events were selected on the 
basis of their shape and orientation defined in a single parameter, azimuthal 
width ({\it azwidth}) (Vacanti et al. \cite{vacanti91}). 
During the season 1988-1989 this instrument was used to detect steady TeV 
emission from the Crab Nebula at the 20$\sigma$ level. 
No evidence was found by O'Flaherty et al. (\cite{o'flaherty92}) for steady, 
modulated, or pulsed emission from Cygnus X-3.
A two-dimensional analysis of the field of view was not carried out at the 
time. 

After each night's observations, the data were archived as binary files on 
magnetic tapes for subsequent analysis. 
During the early 1990's the entire 109-pixel camera database was transferred 
to digital tape (DAT). 
These DAT's formed the data archive for the subsequent analysis.
Table~\ref{database} summarises the database used in this analysis. 
Note that the precise list of the data files originally used by O'Flaherty et 
al. (\cite{o'flaherty92}) in the search for emission from Cygnus X-3 was no 
longer available. 

\begin{table}
  \caption[]{Summary of data.}
  \label{database}
  \begin{tabular}{cccccc}
  \hline
  \noalign{\smallskip}
  Year & No. of data pairs & Duration & N(on) & N(off) & S \\
       &                   & (hours)    &       &        & ($\sigma$) \\
  \noalign{\smallskip}
  \hline
  \noalign{\smallskip}
  1989  & 43 & 26.8 & 367,783 & 368,508 & -0.8 \\
  1990  & 40 & 23.5 & 280,759 & 280,904 & -0.2 \\
  Total & 83 & 50.4 & 648,542 & 649,412 & -0.8 \\

  \noalign{\smallskip}
  \hline

  \end{tabular}
\end{table}

In the re-analysis a small amount of the original data recorded in 1988 
was omitted due to the elevated levels of electronic noise present. 
Slightly more data for the years 1989 and 1990 was included in the analysis. 
This initial selection of data was made prior to the search for gamma-ray
sources. 
The average elevation of the observations was about 70$^\circ$. 
The statistical significance (S) of any excess of events seen in the ON region
is calculated using the combined uncertainty in the ON and OFF event numbers
(Li \& Ma \cite{lima83}).
There is no evidence for an excess of events from Cygnus X-3 in the unselected 
(raw) data.

\begin{table}
  \caption{{\it Supercuts} selection criteria.}
  \label{supercuts}
  \begin{tabular}{ll}
  \hline
  \noalign{\smallskip}
  Trigger & 2 phototubes $>$ 40 digital counts\\
  \noalign{\smallskip}
  Shape   & 0.51$^\circ$ $<$ {\it Distance} $<$ 1.1$^\circ$\\

          & 0.073$^\circ$ $<$ {\it Width} $<$ 0.15$^\circ$\\

          & 0.16$^\circ$  $<$ {\it Length} $<$ 0.30$^\circ$\\
  \noalign{\smallskip}
  Orientation & 0.51$^\circ$  $<$ {\it Distance} $<$ 1.1$^\circ$\\

          & {\it Alpha} $<$ 15$^\circ$\\
  \noalign{\smallskip}
  \hline
  \end{tabular}
\end{table}

\section{Analysis}

As the Cherenkov imaging technique evolved, a more effective
set of gamma-ray image criteria known as {\it supercuts} was developed, 
which separately utilised the shape and orientation information in the image.
The Crab database of 1988-89 was used to optimise the parameters in 
{\it supercuts} (Punch et al. \cite{punch91}) and the criteria were 
subsequently applied to observations of the Crab Nebula from the 1989-90 and 
1990-91 seasons giving significant improvements in sensitivity (Lang et al. 
\cite{lang91}). 
{\it Supercuts} was then adopted as the standard analysis technique for 
observations with this camera (Reynolds et al.\cite{reynolds92}), and, 
with minor modifications, has remained the standard selection method for 
subsequent higher resolution instruments. 
The {\it supercuts} selection criteria are listed in Table~\ref{supercuts}. 
A standard {\it supercuts} analysis of the database was made in which the 
image {\it distance} and {\it alpha} parameters were calculated relative to 
the source position reported by Aharonian et al. (\cite{aharonian02}). 
It was assumed that the optic axis of the camera was centred on Cygnus X-3. 
After de-rotation of the field of view by the parallactic angle, this location 
is displaced from the centre of the field of view by 0.09$^\circ$ in the 
x-axis (corresponding to Right Ascension) and by 0.54$^\circ$ in the y-axis 
(corresponding to Declination). 
The results are listed in Table~\ref{location}.  There is a 3.3$\sigma$ 
excess present at this location. 
As this is an a priori search location there are no trial factors to be 
included and the presence of a weak TeV source is confirmed.

\begin{table}
  \caption{Results for HEGRA location.}
  \label{location}
  \begin{tabular}{lcccc}
  \hline
  \noalign{\smallskip}
   & N(on) & N(off) & N(on)-N(off)& S \\
  \hline
  \noalign{\smallskip}
  Shape             & 12,333 & 12,446 &  -113  & -0.72\\
  Orientation       & 42,607 & 42,692 &   -85  & -0.29\\
  Shape+Orientation &   2031 &   1824 &  +207  & +3.33\\
  \noalign{\smallskip}
  \hline
  \end{tabular}
\end{table}

A two-dimensional search technique has previously been applied to the 
Crab Nebula database of 1988-89 by Akerlof et al. (\cite{akerlof91}). 
The field of view was divided using a Cartesian grid system covering an 
interval of $\pm$1.0$^\circ$ in both the Right Ascension and Declination 
directions. 
The grid step size was 0.1$^\circ$, giving a total of 441 mesh points. 
An {\it azwidth}-type analysis was carried out at each mesh point and 
the Crab Nebula signal appeared as an approximately two dimensional 
Gaussian function with a circular probable error of radius 0.21$^\circ$. 
This is the gamma-ray point spread function (PSF) of the instrument. 
A similar analysis was carried out for observations in which the Crab 
Nebula was deliberately offset by 0.4$^\circ$ and 1.0$^\circ$ from the 
centre of the field of view. 
In each case a signal was detected. 
The efficiency of the telescope (measured in $\sigma$~hr$^{-1}$) was 
undiminished when the source was offset by 0.4$^\circ$ and was reduced to 
45\% for the offset of 1.0$^\circ$. 
At each offset the source location reconstructed from the shower images 
agreed with the tracking location to within approximately 0.03$^\circ$, 
with the dominant error attributed to the telescope angle encoders 
and drive system.

\begin{figure}
  \centering
  \includegraphics[angle=-90,width=8.8cm]{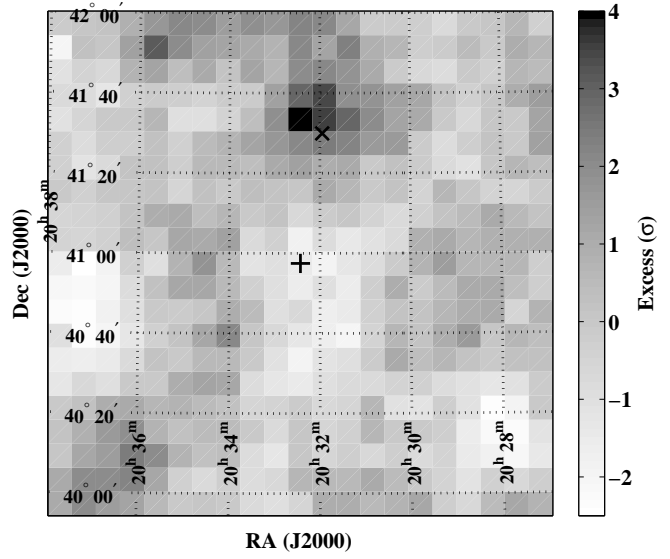}
  \caption{Sky map of the excess significance ($\sigma$) in a 
2$^\circ$x2$^\circ$ region centred on Cygnus X-3 (marked with a +). 
The HEGRA position for TeV~J2032+4130 is marked with an $\times$.}
  \label{nopadmap}
\end{figure}

We applied this two-dimensional analysis to the Cygnus X-3 database, 
where this time events were selected using the {\it supercuts} criteria at the 
441 mesh points. 
Figure~\ref{nopadmap} is a sky map of the ON minus OFF signal over the 
grid. 
The significance of the pre-trial peak signal is 4.0$\sigma$, corresponding 
to an excess of 242 events. 
The peak occurs at the grid point 0.0$^\circ$ in the Right Ascension 
direction and +0.6$^\circ$ in  Declination, 
(RA$_{J2000}$~=~ 20h32m, Dec$_{J2000}$~=~+41$^\circ$33').
which is compatible with both the Crimean and HEGRA positions.
As a conservative estimate of the number of trial factors introduced by 
searching ``near'' the HEGRA position we have used the value 7, which 
corresponds to a search over one quarter of the field of view. 
This reduces the significance of the peak to 3.5$\sigma$. 
If a linear fall-off in sensitivity with displacement over the interval of 
0.4$^\circ$ to 0.6$^\circ$ from the centre of the field of view is assumed, 
the gamma-ray rate corresponds to 12\% of the Crab flux, 
(5.6$\pm$1.6$_{stat}$$\pm$0.8$_{sys}$) x10$^{-12}$ 
photons cm$^{-2}$ s$^{-1}$, 
above the peak energy response of 0.6~TeV. 
The systematic error is representative of the uncertainty in the off-centre
sensitivity of the telescope.

There is no evidence that the signal is variable over the two years of 
observation.
Given the gamma-ray PSF value and the weak nature of the signal, we are unable 
to determine the angular size of the source.

Although the signal is too weak to allow the determination of an energy 
spectrum we have investigated the effect of raising the energy threshold in 
software. 
The image parameter {\it size} measures the total number of photoelectrons in a
Cherenkov image, and therefore the gamma-ray energy.
Table~\ref{sizecut} lists the results when the data is selected on the basis 
of image {\it size}. The cutoff point of 600 digital counts corresponds to a 
doubling of the peak energy response to about 1.2~TeV. 
It would appear that the signal is not confined to low energy events.

\begin{table}
  \caption{Peak signal results selected by image {\it size} (pre-trials).}
  \label{sizecut}
  \begin{tabular}{lcccc}
  \hline
  \noalign{\smallskip}
   & N(on) & N(off) & N(on)-N(off)& S \\
  \hline
  \noalign{\smallskip}
  Size $\le$ 600 dc & 1,574 & 1,405 & 169 & +3.10\\
  Size $>$ 600 dc     &   395 &   322 &  73 & +2.73\\
  \noalign{\smallskip}
  \hline
  \end{tabular}
\end{table}

\begin{figure}
  \centering
  \includegraphics[angle=-90,width=8.8cm]{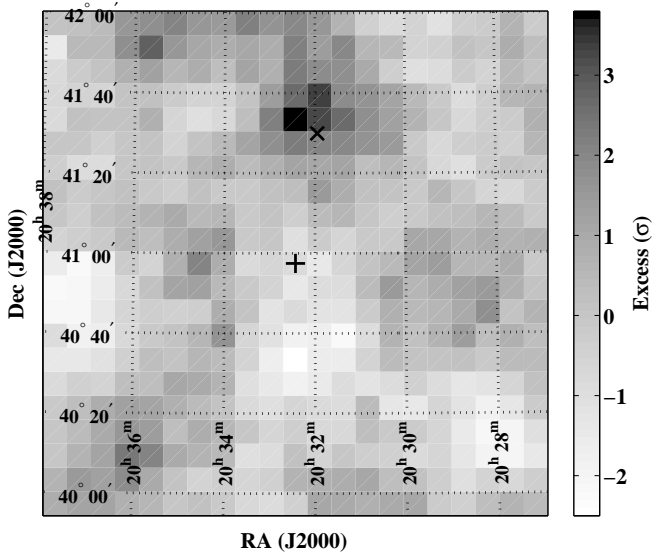}
 \caption{Sky map of the excess significance ($\sigma$) in a 
2$^\circ$x2$^\circ$ region centred on Cygnus X-3 (marked with a +),
after the application of software padding.  
The HEGRA position for TeV~J2032+4130 is marked with an $\times$.}
  \label{padmap}
\end{figure}

The use of software padding procedures, in which Gaussian noise is added in 
order to equalise the night-sky noise in the ON and OFF regions, was not part 
of the standard analysis technique for the 109-pixel camera when this data
was taken. 
Padding was only used in cases of ``extreme'' brightness differences (Punch
et al. \cite{punch91}). 
The Cygnus X-3 ON and OFF regions have similar sky brightness. The night-sky 
noise in each phototube typically had a standard deviation of 3$\sim$4 
photoelectrons, with the difference between the ON and OFF regions having an 
average value of 0.05 photoelectrons.
However, given the weakness of the signal detected from TeV~J2032+4130, the 
effect of software padding on the database was investigated. 
Where $\sigma_{ON}$ and $\sigma_{OFF}$ are the standard deviations of the sky 
noise in a particular phototube in the ON and OFF regions respectively,  
the standard deviation of noise to be added in the less noisy region is given 
by\\
\begin{equation}
  \sigma_{add} = \sqrt{|\sigma_{ON}^2 - \sigma_{OFF}^2|}.
\end{equation}
Software padding is discussed in detail  by Lessard et al. (\cite{lessard}). 
The two-dimensional analysis was repeated ten times, where in each case  
software padding using a different seed for the random number generator was 
applied. 
In order to more closely reproduce the effect of real night-sky noise on the 
event trigger rate, the trigger threshold was raised to two phototubes 
exceeding  45 digital counts. 
This creates a population of sub-threshold events which may trigger 
the system after the addition of noise.
In each case a peak was found at the HEGRA position. 
Because a random feature was introduced into the analysis, the pre-trials peak 
significance varies between 3.1$\sigma$ and 4.5$\sigma$. 
To produce a robust padded analysis the average number of events from the ten 
analyzes which pass the {\it supercuts} criteria at each mesh point was 
calculated. 
These averages were then used to determine the ON minus OFF significance at 
each mesh point. 
The resulting sky map is illustrated in Figure~\ref{padmap}, and is very 
similar to the sky map for the unpadded data, confirming that the effect seen 
does not result from a difference is night-sky brightness. 
The significance of the pre-trials peak is 3.8$\sigma$. 
Given that some noise has been added and the trigger threshold has been 
increased, this slight reduction in significance is not unexpected. 

A subset of the database, comprising 24 out of 83 data pairs, was recorded 
in the OFF-before-ON mode, where the OFF region was 40 minutes ahead in Right 
Ascension.
The average ON minus OFF pre-trials signal per pair in the OFF-before-ON data 
is (0.39$\pm$0.20)$\sigma$ compared with (0.46$\pm$0.13)$\sigma$ in the 
ON-before-OFF data. 
Again this indicates that the signal is not a sky-noise artifact.

\section{Discussion}

TeV~J2032+4130 is located within the crowded Cygnus region of the Galactic 
Plane and hence it has no shortage of possible candidate identifications. The
region is dominated by the very strong OB association, Cygnus OB2 and this 
was the first identification suggested (Aharonian et al. \cite{aharonian02}). 
At a distance of 1.7kpc, this implies a luminosity in the range 10$^{32-34}$ 
erg which is a small fraction of the total energy estimated for 
the OB association. The apparent finite size of the TeV source would
favour this identification.

An extensive study of the source region by Butt et al. (\cite{butt03}) 
using the Chandra and VLA telescopes was made to examine the hypothesis that 
the large scale shocks and turbulence induced by multiple interacting stellar 
winds from the young stars in the OB association might accelerate the hadronic 
component of the cosmic radiation and cause the emission of gamma rays. 
It was assumed that the TeV source was extended and steady. This work has been 
extended by Torres et al. (\cite{torres04}) who predict that this may be a 
strong source of TeV neutrinos.

An alternative approach was taken by Mukherjee et al. (\cite{mukherjee03}) who 
associated TeV~J2032+4130 with the nearby EGRET unidentified source 
3EG~J2033+4118 and assumed the TeV source to be variable and point-like. 
They suggested an identification with the brightest X-ray source in the field
which is variable and has a featureless spectrum suggestive of a BL Lac. 
This was suggested to be a ``proton blazar'' which would be notable for having 
little radio emission.

Other possible identifications include an unusual supernova remnant 
caused by a supernova some tens of thousands of years ago (Bednarek 
\cite{bednarek04}) and the jet from the nearby 
microquasar, Cygnus X-3 (Aharonian et al. \cite{aharonian02}).
 
The detection of a relatively weak signal at a somewhat lower peak energy 
and at an earlier epoch is significant in that it verifies that the source
is variable and hence aids in its identification with candidate objects. The
detection is not deep enough to improve the positional information
or to better define the energy spectrum. Also it is not possible to determine 
the angular size of the source which, like the other TeV sources detected by 
these experiments, is consistent with a point source. 
However the average flux detected in 1989-1990 (12\% of the Crab) is clearly 
well above that seen as the average flux over a four year period by the HEGRA 
group (3\% of the Crab) in 1999-2002 and well below that seen over a six week 
period in 1993 by the Crimean experiment (approximately the level of the 
Crab). 
The low level of emission from TeV~J2032+4130 in the current epoch is 
confirmed by observations at the Whipple Observatory in 2003 
(Finley \cite{finley03}). 
Neither the Whipple nor the HEGRA experiments see any evidence for variability 
within their individual databases. 
The large differences between the flux levels cannot be explained as errors in 
estimation of the sensitivity of the three experiments since they have been 
calibrated by the simultaneous observations of other TeV sources. 

The observed variability is difficult to reconcile with the HEGRA 
observation that the source is extended. It is unfortunate that the
two early experiments did not have sufficient depth in their detections 
to confirm this extension. The variability seen is easier to explain in 
terms of a point source such as the proton blazer or the microquasar
explanations. More sensitive observations by VERITAS and GLAST 
over the full energy range will probably be needed to unambiguously
unravel the nature of the source. The correlation of time variations with
observations at longer wavelengths will be particularly important.

\begin{acknowledgements}
The contributions of  M.F. Cawley, K.S. O'Flaherty and G. Vacanti to the 
observational programme, and to the development of the original analysis 
software, are gratefully acknowledged.
\end{acknowledgements}

\end{document}